# Finite Large Antenna Arrays for Massive MIMO: Characterization and System Impact

Cheng-Ming Chen, *Student Member, IEEE*, Vladimir Volski, *Member, IEEE*,
Liesbet Van der Perre, *Member, IEEE*, Guy A. E. Vandenbosch, *Fellow, IEEE*,
and Sofie Pollin, *Senior Member, IEEE*

*Abstract*—**Massive MIMO is considered a key technology for 5G. Various studies analyze the impact of the number of antennas, relying on channel properties only and assuming uniform antenna gains in very large arrays. In this paper, we investigate the impact of mutual coupling and edge effects on the gain pattern variation in the array. Our analysis focuses on the comparison of patch antennas versus dipoles, representative for the antennas typically used in massive MIMO experiments today. Through simulations and measurements, we show that the finite patch array has a lower gain pattern variation compared with a dipole array. The impact of a large gain pattern variation on the massive MIMO system is that not all antennas contribute equally for all users, and the effective number of antennas seen for a single user is reduced. We show that the effect of this at system level is a decreased rate for all users for the zero-forcing MIMO detector, up to 20% for the patch array and 35% for the dipole array. The maximum ratio combining on the other hand, introduces user unfairness.**

*Index Terms*—**Antenna array mutual coupling, antenna measurements, antenna radiation patterns.**

## I. Introduction

MASSIVE MIMO proposes a new wireless communication concept relying on an excess number of base-station (BS) antennas, relative to the number of active user terminals. The technique allows for very efficient spatial multiplexing, attainable using linear processing in a time-division duplex mode [1]–[3]. It has been demonstrated to achieve a record spectral efficiency (SE) [4]. Moreover, the technology has the potential to drastically improve energy efficiency [5]. Consequently, massive MIMO addresses several key 5G requirements [6]: it offers a great capacity increase, can support more users, and enables significant improvement in energy efficiency.

Massive MIMO operation has been studied extensively relying on omnidirectional profiles and homogeneous

Manuscript received December 30, 2016; revised June 27, 2017; accepted July 25, 2017. Date of publication September 19, 2017; date of current version November 30, 2017. This work was supported in part by the European Union Seventh Framework Program (FP7/2007-2013) through MAMMOET under Grant 619086 and in part by the Flemish Hercules Foundation under Grant AKUL1318. *(Corresponding author: Cheng-Ming Chen.)*

The authors are with the ESAT-TELEMIC Research Division, Department of Electrical Engineering, Katholieke Universiteit Leuven, 3001 Leuven, Belgium (e-mail: cchen@esat.kuleuven.be; volski@esat.kuleuven.be; lvanderp@esat.kuleuven.be; vdbosch@esat.kuleuven.be; spollin@esat.kuleuven.be).

Color versions of one or more of the figures in this paper are available online at http://ieeexplore.ieee.org.

Digital Object Identifier 10.1109/TAP.2017.2754444

arrays [7]–[10]. Most of these studies neglect the impact of the directional array gain pattern on the massive MIMO system performance. These assumptions are overoptimistic for realistic scenarios with compact antenna arrays. These feature a finite-number of antennas that are spaced relatively close to each other (a typical example is half a wavelength) and hence can experience significant mutual coupling. In [13] and [20], the analytic massive MIMO sum-rates has taken mutual coupling into account and show that channel correlation is dependent on mutual impedance. However, the mutual impedance was derived from single element, so there is no gain pattern variation in the considered model. It has also been shown that in most realistic scenarios the channels deviate from the i.i.d. Rayleigh assumption, and the gain variation of the channels impacts the overall system capacity because not all antennas contribute equally [8]. These studies were either on a virtual array, neglecting the mutual coupling, or study the gain variation combined with the multipath channel. The impact of gain variations caused purely by the antenna array is not yet studied. Moreover, measuring the impact of the array topology on the active or embedded gain pattern of a single element requires an antenna measurement facility where multiple antennas can be active at the same time. Most antenna measurements create a virtual array, by moving the antenna along a plane [9], [10] or measure antennas in an array where only a subset of antennas are active at the same time [10]–[12]. Active array antenna measurements have to the best of our knowledge not yet been reported.

The realized gain of a single antenna is a very important parameter. Typically, it is one of the parameters specified in the datasheets. However, in the case of arrays, the realized gain of identical elements can significantly vary due to the mutual coupling, or in other words the electromagnetic interaction between elements. Mutual coupling is a changing of currents in one element, which creates a field that changes the currents on adjacent elements. Hence, this changes the realized gain of each antenna element. These parasitic-induced currents affect all parameters of the elements: S-parameters and embedded gains. So the description of mutual coupling based on S-parameters is related to power flows between the elements, while embedded gain patterns also involve the directions in space where the power radiates. The latter depends strongly on induced currents and on the type of interference: constructive or destructive [18], [19].

In the existing literature [2], [3], there are clear no guide-lines of how to select a basic element for a massive MIMO antenna array, although this is really a crucial aspect of a massive MIMO array and system. One thing that is known from the basic MIMO theory is that it is always better if an antenna element in such an array receives as much multipath from all directions as possible. Hence, it has often been assumed that using a quasi-omnidirectional dipole is always better than the more directive patch element.

In this paper, for the first time, the effect of mutual coupling in larger arrays on embedded gains, and specific the conse-quent impact on the system performance in a massive MIMO system is investigated, both for the more omnidirectional dipole element, and the well-known and widely used patch element. This is done by, including the gain variation into the small-scale fading channel model. The study of how these realized gain variations (a problem more understood in the antenna and propagation community) impact system-level performance (a problem formulation approach typically used in the massive MIMO signal processing community) is novel and of great interest to both communities.

We first study the active gain pattern variation of individual antenna elements in a large massive MIMO array, caused by the mutual coupling between the closely located elements and the edge effects in finite arrays. Both dipoles and patch antennas are considered in the simulation-based assessment, and for the latter results of real-life experiments are also presented. Our antenna measurements rely on measuring 32 active elements in an array, which is enabled by relying on a massive MIMO test bed placed in an anechoic chamber. Consequently, the impact of the gain pattern variation on the achievable SE is highlighted. While a dipole individually features a better omni-directionality, when composed in an array their severe mutual coupling causes drastic directionality on the elements and gain variations over the array. The patch array is shown to be the better choice from the system capacity point of view.

This paper is further organized as follows. First, we intro-duce a massive MIMO system model with an extended chan-nel model that takes into account the 3-D antenna gain in Section II. Next, the simulation-based assessment of antenna gain variation and directivity of a representative finite large array composed of either dipoles or patch antennas is provided in Section III. The experimental validation is presented in Section IV. The impact of the gain variation on SE at system level is illustrated in Section V. Finally, we conclude this paper by reviewing the main findings, and provide recommendations for the design of large antenna arrays to be used in massive MIMO systems.

The notation used in this paper is as follows: We denote bold face upper (lower) letters as matrices (vectors). Super-scripts $H$, $T$, and $-1$ stand for Hermitian transpose, trans-pose, and inverse, respectively. The matrix $\boldsymbol{I}_K$ denotes an $K \times K$ identity matrix. Moreover, $\otimes$ denotes as Kronecker product, vec{.} represents vectorization of a matrix, det(.) is the determinant of a matrix, and cofactor(.) means the cofactor operation of a matrix. The element in the $k_t h$ row and $m_t h$ column of matrix $A$ is denoted by $[A]_{k,m}$.

## II. 3-D SYSTEM MODEL

In this section, we introduce the system model bringing into account a 3-D gain pattern for the antenna elements in the array. The actual 3-D gain pattern at each antenna element depends both on the embedded gain pattern, as well as the various multipath reflections. This requires the establishment of a fairly detailed channel model, including propagation and array gain patterns. To access the impact from the gain variation to the system performance, we later plug the results of arrays consisting of dipoles or patch antennas in Section IV into this channel model and simulate the impact of gain variation to the user achievable rate in Section V-B.

A massive MIMO BS equipped with $M$ antennas communi-cates with $K$ single-antenna user terminals in the same time–frequency unit. The symbols transmitted from the $K$ users are represented as a vector $s = [s_1, \dots, s_K]^T$, where $\mathbb{E}[|s_k|^2] = 1$. The received signal $y$ after transmission over the channel and disturbance by noise is

$$y = \boldsymbol{D} \boldsymbol{X}^{1/2} s + \boldsymbol{w} \tag{1}$$

where $y \in \mathbb{C}^M$, $X = \text{diag}\{x_1, \dots, x_K\}$ with $x_k$ denoting the average transmit power of user $k$, while $\boldsymbol{w} \sim \mathcal{CN}(\boldsymbol{0}, \boldsymbol{I}_M)$ is the i.i.d. complex Gaussian distributed noise. $D = [\boldsymbol{d}_1, \dots, \boldsymbol{d}_K]$ represents the channel, with the channel vector between the M-antenna BS and the $k_{th}$ user $\boldsymbol{d}_k \in \mathbb{C}^M$. Originating from the correlation channel model in [14], we decompose the channel vector $\boldsymbol{d}_k$ into three terms, namely, large-scale fading, antenna gain variation, and small-scale fading

$$\boldsymbol{d}_k = \frac{\sqrt{\alpha_k}}{C_k} \sum_{c=1}^{C_k} \boldsymbol{G}(\theta_{c,k}, \phi_{c,k}) \Delta_c \boldsymbol{a}(\theta_{c,k}, \phi_{c,k}) v_{c,k} \tag{2}$$

where $\alpha_k$ represents the large-scale fading and shadowing effect of user $k$ seen by the whole antenna array and $C_k$ stands for the number of multipath components. The array gain pattern is a diagonal matrix $\boldsymbol{G}(\theta_{c,k}, \phi_{c,k}) = \text{diag}\{(g_1(\theta_{c,k}, \phi_{c,k}))^{1/2}, \dots, (g_M(\theta_{c,k}, \phi_{c,k}))^{1/2}\}$, which repre-sents the different active antenna patterns from different angle of arrivals for each antenna $m$ due to mutual coupling and the edge effect. To represent the rich multipath environment, $\Delta_c$ is an $M \times M$ matrix with binary diagonal elements

$$[\Delta_c]_{m,m} = \begin{cases} 1, & \text{belongs to cluster} \\ 0, & \text{otherwise} \end{cases} \tag{3}$$

specifying whether the reflection belongs to the multipath cluster $c$. This matches the fact that for a large antenna array, reflections from one cluster do not contribute to all antennas. The steering vector $\boldsymbol{a}(\theta_k, \phi_k)$ of a rectangular matrix is modeled as

$$\begin{aligned} \boldsymbol{a}(\theta_k, \phi_k) = \text{vec} \Big\{ &\Big[ 1, e^{j2\pi \frac{\gamma}{\lambda} \sin \theta_k}, \dots, e^{j2\pi (\sqrt{M}-1) \frac{\gamma}{\lambda} \sin \theta_k} \Big]^T \\ &\otimes \Big[ 1, e^{j2\pi \frac{\gamma}{\lambda} \sin \phi_k}, \dots, e^{j2\pi (\sqrt{M}-1) \frac{\gamma}{\lambda} \sin \phi_k} \Big] \Big\} \end{aligned} \tag{4}$$

where $\gamma$ is the antenna spacing, $\lambda$ is the carrier wavelength, and $\phi_k$ denotes a azimuth of arrival angle. Moreover, $v_{c,k} \sim \mathcal{CN}(0, I)$ represents a standard complex

Gaussian vector. When there is only a single line-of-sight (LoS) cluster, the model simplifies to

$$\mathbf{d}_k = \sqrt{a_k}\mathbf{G}(\theta_k, \phi_k)\mathbf{a}(\theta_k, \phi_k). \qquad (5)$$

For the simulation results in Section V, we use the simplified channel model in (5) to consider the effect of pure antenna patterns. However, we develop a more general channel model in (2) illustrating that the assessment of system-level impact of gain variations is not trivial.

## III. Gain Pattern in Large Arrays: Dipoles Versus Patch Antennas

It is favorable for each antenna element in massive MIMO to have equal gain from all directions so as to efficiently exploit the multipath in the wireless environment. Typically, researchers assume an antenna element that preserves its characteristics in an array environment [9], [10]. However, in practice the mutual coupling between closely spaced elements may noticeably affect the embedded element radiation pattern, making it different from the pattern of a single element.

An accurate computational analysis of such influence requires a full wave solver, which is capable of taking into account the mutual coupling between elements and is able to calculate the embedded gain pattern of each element. In this paper, CST microwave studio has been used to compare the gain patterns of a single-antenna element, a finite array, and an infinite phased array. Since it is of interest to compare the qualitative performance of different types of antenna element, a more directional and a more omnidirectional antenna element have been considered. The first type is a microstrip patch antenna and the second type is a half wavelength dipole that generates an omnidirectional pattern in the H-plane.

The microstrip patch prototype consists of a square patch of 31 mm with two merged U-slots with width 1.4 mm. Then, the patch and slot shapes were deformed to polygons using the optimization procedure in CST to cover the frequency bands 2.4–2.62 and 3.4–3.6 GHz. The main comparison in this paper has been performed at 2.6 GHz. A single patch is shown in Fig. 1. The patch is etched on a 1.6 mm FR4 substrate mounted on 5 mm nylon spacers above another 1.6 mm FR4 substrate. The antenna dimensions are 70 × 70 mm. The dimension of dipole is about 51.3 × 2 mm. Both types of finite arrays are illustrated in Fig. 2,[1] with an element spacing of 71 mm.

A first estimation of mutual coupling can be obtained from the analysis of the simulated S-parameters as shown in Fig. 3 for the elements in the center and in the corner. All elements in the array are consecutively numbered from the left bottom corner as shown in Fig. 2. The simulated reflection coefficient for a single element are also plotted with curves labeled single in superscript. The simulated mutual coupling between the dipoles in Fig. 3(b) is higher in comparison with the simulated mutual coupling between patch antennas in Fig. 3(a) by around 6 dB. Furthermore, in order to illustrate the accuracy of these simulations, representative measurements were performed in an anechoic chamber using a spectrum analyzer Keysight N9344C with a tracking generator; a typical agreement is illustrated in Fig. 3(a) for $s_{27,28}$.

[1] The spherical coordinate system used in the paper is based on the convention accepted in physics and in the antenna community. The theta angle is counted from the z-axis. The Cartesian coordinate system is defined in Fig. 2.

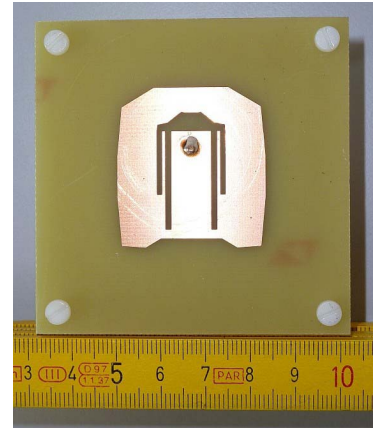

Fig. 1. Detailed view of the microstrip patch antenna.

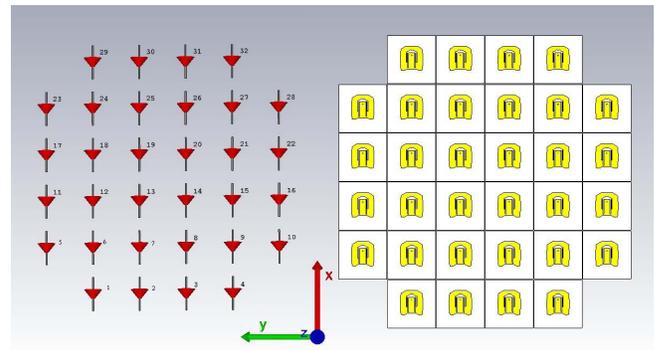

Fig. 2. Two finite 32-element antenna arrays: dipoles (left) and patches (right).

Consider the $k_t h$ user and a single element in the BS in a LoS scenario. The power $p_k^{(r)}$ received by the element can be estimated using the well-known Friis transmission formula

$$p_k^{(r)} = p_k^{(t)} g_k^{(t)} r_k g_k^{(r)} \qquad (6)$$

where $p_k^{(t)}$ is the transmit power from the user and $g_k^{(t)}$ is its realized gain. $g_k^{(r)}$ is the embedded realized gain or active gain pattern of the element in the BS, and $r_k = (\lambda/(4\pi \Delta r_k))^2$ is the inverse of free-space pathloss with distance $\Delta r_k$ between the $kth$ transmitter and the element.

As for an array, the variation in the received power per element is coupled with the embedded gain variation of the elements, so from now on we will focus only on the receive realized embedded gain. For simplicity, the superscript $(r)$ is omitted. For an infinite array, the embedded gain is identical for all elements and can be easily calculated. The calculation reduces to the analysis of a unit cell taking into account a phase shift between neigboring elements. This phase shift depends on the main Floquet harmonic in the direction $(\theta_k, \phi_k)$. The embedded realized gain $g_{m,k}^\infty$ in the infinite array of $m$th element is modulated by the reflection coefficient $\Gamma^\infty$ [15].

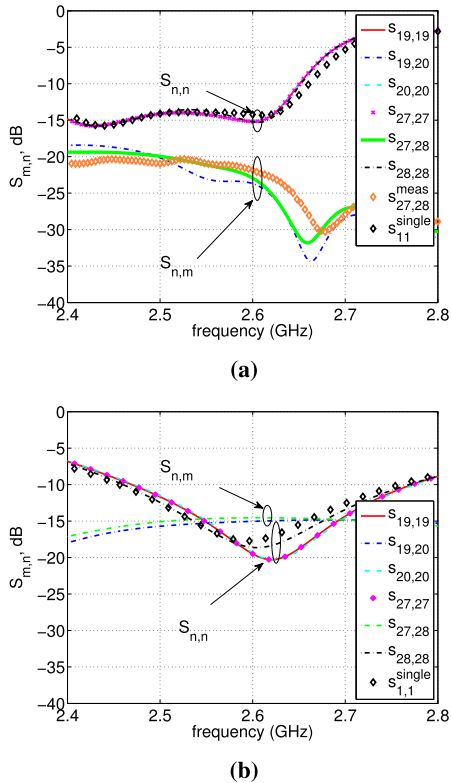

**(a)**

**(b)**

Fig. 3. Mutual coupling between elements selected in the array center and at the edges. (a) Patch array. (b) Dipole array.

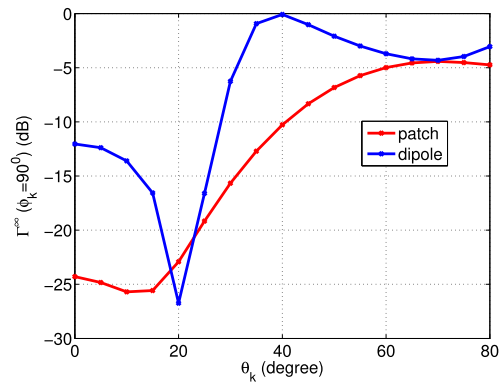

Fig. 4. Reflection coefficient for elements in an infinite phased array at 2.6 GHz.

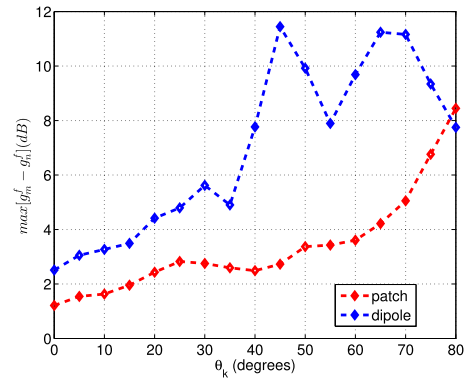

Fig. 5. Maximal gain variation between two elements in terms of direction of incidence in the azimuthal plane of the 32-element finite array at 2.6 GHz.

The simulation result is shown in Fig. 4.[2] When the reflection coefficient goes to 1 for some direction(s), the embedded realized gain goes to zero. These directions are called scan blindness angles (SBAs). Note that in practice, the reflection at SBA can be smaller than 1 due to the losses in dielectric and metal of the antenna elements. The far-field components can be obtained by analyzing the transmission from the antenna port to the main Floquet harmonic. One of the obvious conclusions of this paper is that a strong mutual coupling between elements can completely destroy the omnidirectional pattern of the dipole.

In a finite array, the situation is quite different. There, because of the edge effect, i.e., the fact that the elements at the edges see a different environment compared with the elements in the middle, the embedded gains of the elements are not identical. In this paper, the maximum gain variation over the elements was obtained in three steps. First, for each direction of incidence $(\theta_k, \phi_k)$, the embedded gains of all elements $\boldsymbol{g}_k^f$ are calculated, where the superscript $f$ stands for finite array. Second, for a given $\theta_k$ and $\phi_k$, the maximum difference between two embedded element realized gains is calculated over the whole array $\max_{m,n}(g_{m,k}^f(\theta_k, \phi_k) - g_{n,k}^f(\theta_k, \phi_k))$. Finally, this maximum difference can be studied as a function of direction as depicted in Fig. 5. Two very important observations can be made. First, the maximum gain variation increases considerably when the angle $\theta_k$ approaches the SBA. Second, the patch array shows a lower gain variation between elements at angles closer to the direction normal to the array. This means that, counter-intuitively, the more directive patch elements are the better choice from the point of view of gain variation. In order to study the dynamic range of the array, for each angle $\theta_k$, we plotted $\max_{m,\phi_k}(g_{m,k}^f(\theta_k, \phi_k))$, $\min_{m,\phi_k}(g_{m,k}^f(\theta_k, \phi_k))$, and $\text{mean}_{m,\phi_k}(g_{m,k}^f(\theta_k, \phi_k))$ of the embedded gains in Fig. 6. It is clearly proven that the role of mutual coupling is very destructive: elements that are intrinsically omnidirectional when isolated do not provide an omnidirectional coverage any more in the finite array environment. As long as $\theta_k$ is less than 60°, the dynamic range of the patch element is around 5 dB, which is 5 dB less than that of the dipole array.

## IV. MEASURED ACTIVE GAIN PATTERNS

In order to validate the active gain variations predicted by the simulations, measurements were performed on the finite 32-element patch array in receive. The operating frequency was 2.6 GHz and, obviously, the element distance was 71 mm. Both the patch array and a wide-band horn (EMCO 3115) transmit antenna were located inside the anechoic chamber at KU Leuven with 7 m of distance in between, as shown in Fig. 7. The patch array was fixed on a cylindrical holder mounted on a positioner capable of rotating in the azimuthal

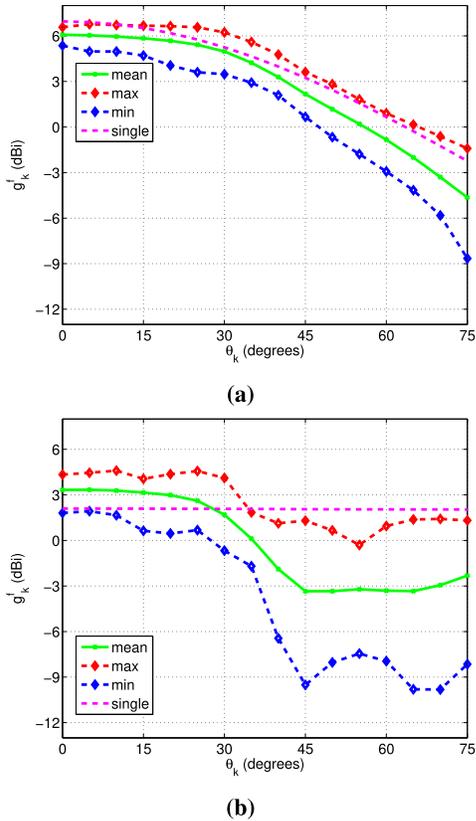

**(a)**

**(b)**

Fig. 6. Embedded realized gain variation for a 32-element finite array. There is a higher gain variation for the dipole array even at angles $\theta_k$ close to the direction normal to the array at 2.6 GHz. (a) Patch array. (b) Dipole array.

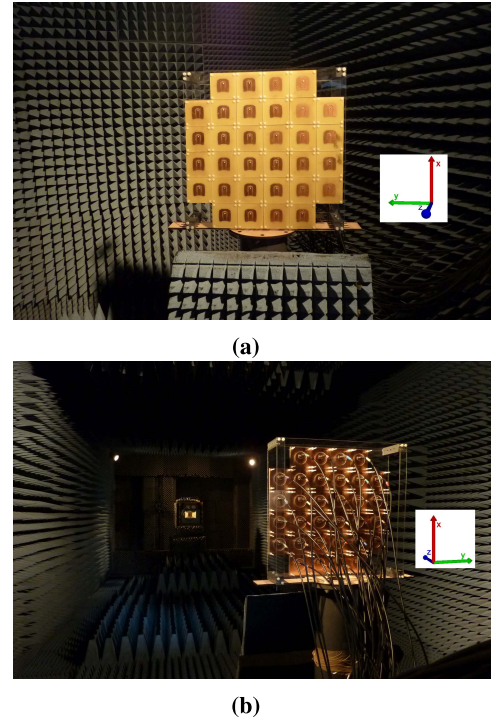

**(a)**

**(b)**

Fig. 7. Measurement setup. (a) 32-element patch array on a round table rotating in the range $\theta_k = -75° : 5° : 75°$. (b) Back view of the antenna array. Both horn and array are in each other's broadside direction when $\theta_k$ equals 0°.

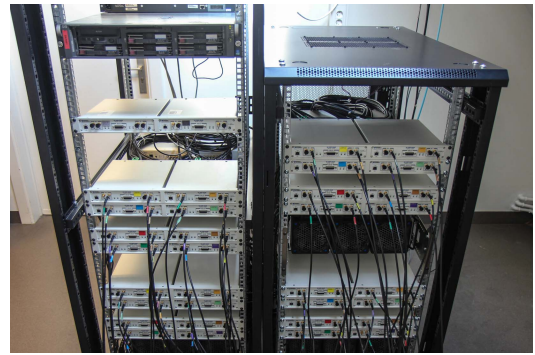

Fig. 8. Measurement setup: power gain variations across the array were calculated from LTE-like uplink data symbols in the KUL massive MIMO test bed. 32 antennas were used in this measurement.

plane. Each patch was connected via 18 m RF cables to MIMO test bed outputs.[3]

The dimension of the patch array is about $44 \times 44$ cm. Following the horn specification, the 3-D beamwidth in the E-plane is of 53° and 48° in the H-plane. So the array illumination should remain relatively uniform and the incident field variation is considerably smaller in comparison with the variation of measured power levels between elements. So all observed variations in the received power levels can be attributed to mutual coupling between antenna elements.[4] The dependence of the gain variation on the angle was validated by performing measurements in the following zenith angles $-75° : 5° : 75°$ (31 discrete angles in the $y$–$z$ plane) while fixing the azimuth angle $\phi_k$ to 90°. Note that, while assuming a thermal noise level of $-174$ dBm/Hz, the SNR of this measurement was above 50 dB. Details of the RF settings are given in Table I.

The synchronized power measurement from 32 antennas was accomplished by a massive MIMO system termed MIMO framework [16] running in the KU Leuven (KUL) MaMi test bed. From which 16 (2 RF ports each) universal software radio peripherals (USRPs) jointed together as a BS as shown in Fig. 8.

For the user side, a single USRP was connected to the horn antenna as a transmitter. The received power strength of the 32-element was calculated from the uplink data symbols synchronized by an LTE-like frame structure.

At each $\theta_k$, 30 s of signal strength were recorded and the statistics of maximum, minimum, and mean from 32 antennas were plotted in Fig. 9. We observe that there is a high power gain variation among the antenna array while the zenith angle deviates from 0°. In addition, the measurements agree with the CST simulation in several aspects. First, the received gain is quite flat when $|\theta_k| \leq \pm 20°$ and within this region, there is a low variation of around 3 dB. Second, the maximum received gain decreases noticeably for larger zenith angles while the gain variation is increasing. The measured gain range at each incident angle is summarized in Fig. 10, which follows the simulation trend with a higher level of about 1 dB. The higher level can be explained by the presence of various supporting elements located in the array environment that were not taken into account during the simulation. To see how the gain variation distributed along the panel with related to different angle of arrivals, we further map the measured gain of each element with its position on the panel at zenith angles 40° and −40° for both simulation and measurement. The received power were normalized to the mean power and shown in Fig. 11. Again, the simulation results are perfectly symmetric for both angles. In addition, the measurement result at $\theta_k = 40°$ matches the simulation quite well over the whole map. For the angle at $\theta_k = −40°$, our measurements show larger deviation from the simulation, which is caused by multipath reflections caused by our openings in the anechoic chamber, as well as induced currents on the RF cables. We observe a larger gain variation in the edge elements compared with the center elements, this is the edge effect. It is very important to note that different elements are sensitive to very

---

[3] The anechoic chamber has an asymmetrical opening for RF cables and the positions of the RF cables are also not ideally symmetrical. Thus this real setup is a little asymmetrical due to several supporting elements leading to a slightly asymmetrical response.

[4] It is also important to remember that the radiation pattern of a patch element in the E-plane is not symmetrical. As a consequence, we do not expect any symmetrical gain measurements in the vertical set of elements for any incident angle.

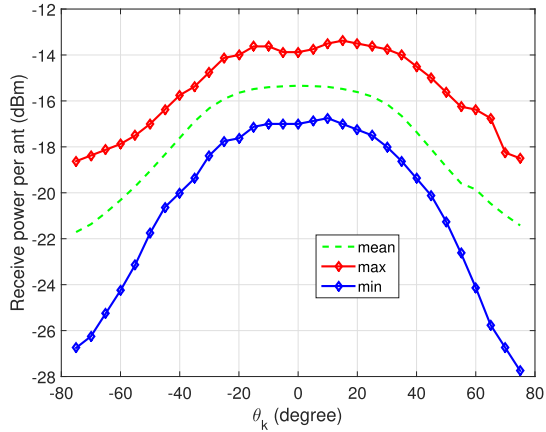

Fig. 9. Measured gain variation by a 32-element rectangular antenna array. The array has a lot of variation at high zenith angles (large difference between max and min).

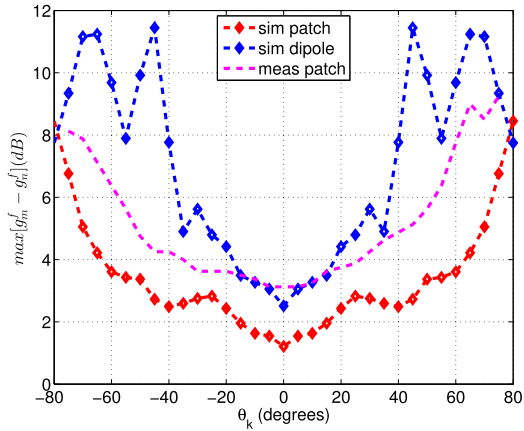

Fig. 10. Maximal measured and simulated gain variation between two elements in terms of direction of incidence in the azimuthal plane at 2.6 GHz.

## TABLE I
### RF Power Settings for Array Gain Measurement

| Parameters | Gain |
|---|---|
| Horn | 8.8 dBi |
| Patch | 6 dBi |
| TX Power | 20dBm |
| RX Gain | 33.5dB |
| Cable Loss | -23.4dB |
| Free Space Path Loss | -57.6 dB |
| Received Level | -12.7dBm |

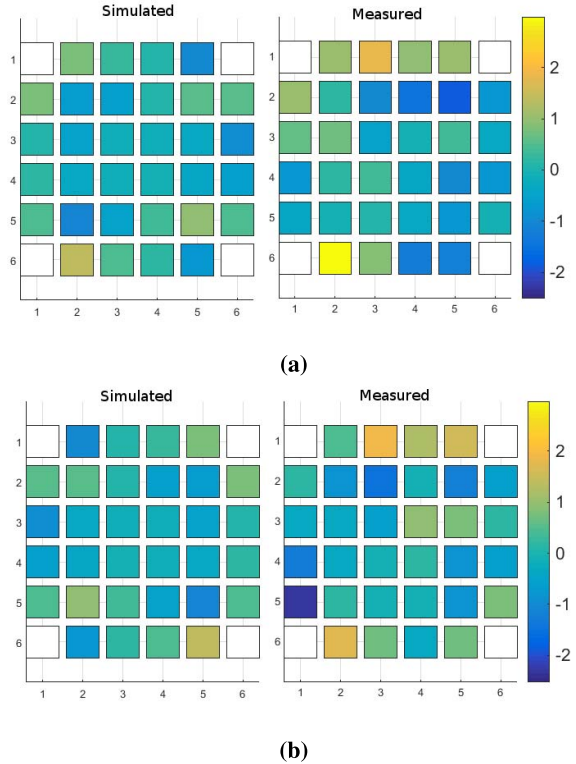

Fig. 11. Illustration of the simulated and measured power when the signal arrives from different angles. It is quite clear to see that received power strength depends on the incident angle. In LoS scenarios, the power received from different users might have different levels of power distribution among the antenna elements. (a) $\theta_k = 40°$. (b) $\theta_k = −40°$.

different directions, i.e., there is a severe gain variation that varies with incident angle. In any case, when the signal comes from different angles, an antenna element that receives a higher power in one direction does not always receive a higher power from the other direction. We should point out that gain variation also increases the required dynamic range for a fixed-point system implementation, as the automatic gain control in the receiver is not capable of jointly optimizing the received power levels from different directions.

## V. Gain Variation and Spectral Efficiency

We have seen that there is a considerable gain variation over the array. Also, there is a different level of gain variation for patch and dipole antenna arrays. In this section, we compare the impact on single user achievable rate in a massive MIMO system. First, to theoretically show how the gain variation affects the single user achievable rate, we introduce the SE metric for both linear maximum ratio combining (MRC) and zero-forcing (ZF) detectors. Then, we apply the measured gain variation from the patch array and the CST simulated gain from the dipole array, respectively, to examine the impact of array pattern variation on a massive MIMO system.

### A. Spectral Efficiency of MIMO Detectors

Under the assumption that the BS has perfect channel state information and the channel is ergodic, the uplink ergodic achievable rate from MRC and ZF detector can be represented as [17]

$$R_k^{mrc} = \mathbb{E}\left\{ \log_2\left(1 + \frac{x_k \|\boldsymbol{d}_k\|^4}{x_k \sum_{i=1, i \neq k}^{K} \|\boldsymbol{d}_k^H \boldsymbol{d}_i\|^2 + \|\boldsymbol{d}_k\|^2}\right)\right\} \quad (7)$$

and

$$R_k^{zf} = \mathbb{E}\left\{ \log_2\left(1 + \frac{x_k}{\|[(\boldsymbol{D}^H\boldsymbol{D})^{-1}]_{k,k}\|}\right)\right\}. \quad (8)$$

The MRC per user rate $R_k^{mrc}$ in (7) illustrates the two main effects that determine the SE of massive MIMO.

1) Due to the array gain, the SNR without considering interuser-interference (IUI) increases linearly with the antenna array size. In our system model, we given the noise power $\sigma_w^2 = 1$, so SNR $= x_k \|\boldsymbol{d}_k\|^4 / \|\boldsymbol{d}_k\|^2$, meaning that is best to have a maximal number of antennas. Antennas with a low gain, do not contribute and reduce the effective number of antennas seen.

2) The user separation enables to spatially multiplex multiple users based on their unique signature at the antenna array. The interuser correlation term $\|\boldsymbol{d}_k^H \boldsymbol{d}_i\|^2$ in the denominator of (7), when considering only two users for simplicity, the IUI term can be represented as

$$\|\boldsymbol{d}_k^H \boldsymbol{d}_i\|^2 = \|\boldsymbol{d}_k\|^2 \|\boldsymbol{d}_i\|^2 \|\cos\theta_{ki}\|^2 \quad (9)$$

where $\cos\theta_{ki}$ is the angle between $\boldsymbol{d}_k$ and $\boldsymbol{d}_i$. Suppose due to gain pattern variation, user $k$ has a higher channel vector two-norm than user $i$. We then obtain the signal to interference ratio (SIR) relationship between user $k$ and $i$ as

$$\text{SIR}_i \leq \text{SIR}_k \iff \frac{\|\boldsymbol{d}_i\|^2}{\|\boldsymbol{d}_k\|^2} \leq \frac{\|\boldsymbol{d}_k\|^2}{\|\boldsymbol{d}_i\|^2}. \quad (10)$$

We call this user unfairness caused by antenna gain pattern variation.

On the other hand, the performance of the ZF detector can be understood by looking into

$$\|[(\boldsymbol{D}^H\boldsymbol{D})^{-1}]_{k,k}\|^{-1} = \frac{\|\det(\boldsymbol{D}^H\boldsymbol{D})\|}{\text{cofactor}(\boldsymbol{D}^H\boldsymbol{D})_{k,k}} \simeq \|\boldsymbol{d}_k\|^2. \quad (11)$$

Here, the Hadamard inequality is applied in the approximation. Hence, we can observe that the achievable rate is directly proportional to the two-norm of the channel vector, including the antenna gain pattern.

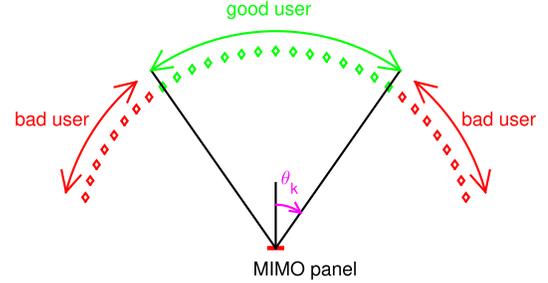

Fig. 12. There are 31 measured locations, where the good user (user one) locates in the region with high power and low gain variation (15 discrete locations), while the bad user (user two) is placed outside this region (16 discrete locations).

### B. Simulated Gain Variation Impact

To simulate the impact of measured antenna gain variation on system SE, we consider a LoS scenario with $M = 32$ and $K = 2$. The two users are assumed to have equal distance to the BS, so we say they share a common large-scale fading $\alpha_k = 1$. Moreover, good user (user one) locates in a higher power and less gain variation region, i.e., in the zenith angles $|\theta_k| \leq 35°$ (15 discrete locations). While a second bad user locates outside this region, i.e., in zenith angles $35° < |\theta_k| \leq 75°$ (16 discrete locations), as illustrated in Fig. 12. Both of their azimuth angles are distributed at a very limited region $\phi_k = 88° : 1 : 92°$. Furthermore, no power control is considered for simplicity, and the transmitted power $x_k$ is assumed to be equal for both users.

We compare the single user achievable rate of both users for the measured patch array and the simulated dipole array. As a patch antenna has higher embedded gain and can be referenced from Fig. 6, the peak power of patch and dipole arrays are normalized to 0 and $-3$ dB, respectively. A reference scenario without gain variation, the peak gain for all angles is set to 0 dB, is also given. Only one user in the no gain variation case is plotted for comparison, as both users have equal performance. First, the per user achievable rate of the MRC detector is plotted as shown in Fig. 13. For each realization, we randomly put one user in the good and one user in the bad region, calculate the rates, and average the two rates over ten thousand realizations. The good user apparently benefits when coexisting with a bad user. A more severe user unfairness is experienced for the dipole array, as the gain pattern variations are more pronounced here. The gain pattern variation increases the rate of good users up to 6% and decreases the rate of bad users up to 24% at an intermediate SNR $= 25$ dB. Second, the ZF achievable rate is shown in Fig. 14. From (11), we see the achievable rate is directly proportional to the received user power and this matches the result that achievable rate of patch is in general higher than that of dipole. If we compare the reference with the bad-power user of dipole array, there is a huge SNR loss by 10 dB and can be improved by 3 dB if instead applying the patch array.

Figs. 13 and 14 are obtained under the assumption that there are always two users actively communicating in the system. The conclusion of the MRC method is that the achievable rates of both users are coupled. The good user causes a larger

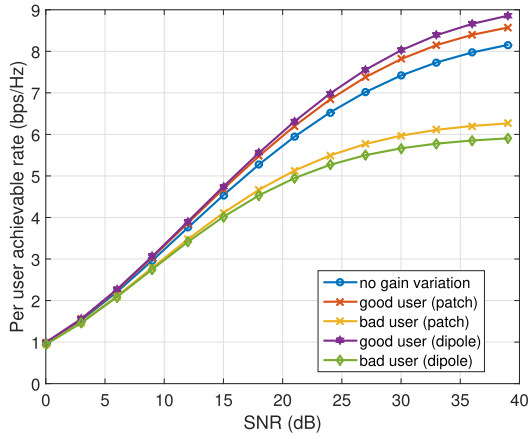

Fig. 13. MRC per user rate. Performance of dipoles exhibits a higher level of user unfairness.

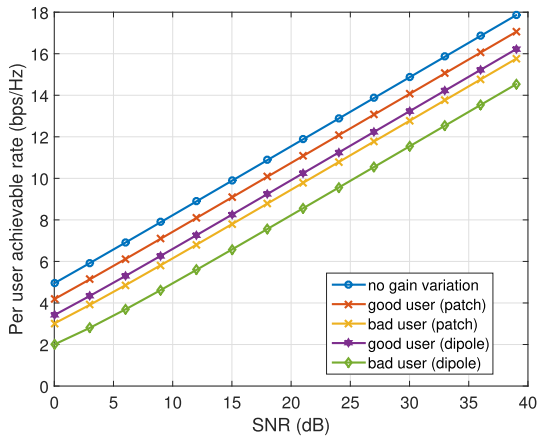

Fig. 14. ZF per user rate. Dipoles counter-intuitively is less omnidirectional. The bad-gain user suffers from lower level of received power hence gets the lowest achievable user rate.

IUI to the bad user which results in a big impact on the achievable rate of the bad user. On the other hand, the bad user induces less IUI, and that is why the achievable rate of the good user is higher than the achievable rate of no gain variation case. We should notice that when there is no gain variation, the two users receive the same peak power from all directions. Moreover, we should highlight that for a fair communication system, all users should receive similar achievable rate instead of some benefits more if the user receives a better channel condition. The performance of each method should be evaluated by the performance of the bad user.

## VI. CONCLUSION

It has often been assumed in theoretical studies on massive MIMO that all antennas contribute equally in a massive MIMO system. In this paper, we experimentally verify that in a finite array, there is a strong variation in the gain pattern of the different antenna elements. This gain pattern variation is caused by mutual coupling and the edge effect, and strongly depends on the angle of arrival. Remarkably, the gain variation is larger in a dipole array, because of stronger mutual coupling in such a system. This makes the array, consisting of omnidirectional elements, more sensitive to angle of arrival than a patch array consisting of directional elements. Because of this angle of arrival-dependent gain variation, the received power over the array is not the same for all the users. While gain variation is potentially beneficial for user separation, the main effect is that the received power from each user is decreased because of suboptimal antenna gains. For the MRC detector, the system-level impact leads to user unfairness as this detector exploits the decreased correlation of the users maximally while disadvantaging the user in a suboptimal angle. For ZF, our assessment shows that all users are disadvantaged by the antenna gain variation, and see a lower rate than a system with ideal identical antennas. Our future work is to investigate appropriate topologies and configurations of the antenna array to reduce the impact of such large gain variation effects.

## ACKNOWLEDGMENT

The authors would like to thank NI Engineer Dr. A. Nasser Ali Gaber for his technical support of the massive MIMO framework.

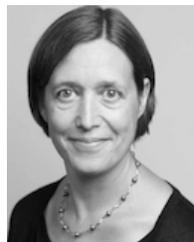

**Liesbet Van der Perre** (M'02) received the Ph.D. degree from the Katholieke Universiteit Leuven, Leuven, Belgium, in 1997.

She was with the Nano-Electronics Research Institute, imec, Leuven, from 1997 to 2015, where she took on responsibilities as a Senior Researcher, a System Architect, Project Leader, and Program Director. She was appointed Honorary Doctor with Lund University, Sweden, in 2015. She was the Scientific Leader of the EU-FP7 Project MAMMOET on Massive MIMO. She is currently a Professor with the Department of Electrical Engineering, Katholieke Universiteit Leuven, and a Guest Professor with the Electrical and Information Technology Department, Lund University, Lund, Sweden. She has been a member of the Board of Directors of Zenitel, Belgium, since 2015. She has authored or co-authored over 300 scientific publications. Her current research interests include wireless communication, with a focus on physical layer and energy efficiency.

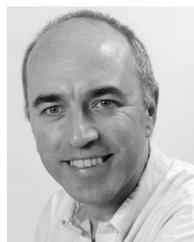

**Guy A. E. Vandenbosch** (F'13) received the M.S. and Ph.D. degrees in electrical engineering from the Katholieke Universiteit Leuven, Leuven, Belgium, in 1985 and 1991, respectively.

Since 2005, he has been a Full Professor with the Katholieke Universiteit Leuven. In 2014, he was a Visiting Professor with Tsinghua University, Beijing, China. He has taught courses on electromagnetic waves, antennas, electromagnetic compatibility, fundamentals of communication and information theory, electrical engineering and electrical energy, and digital steer and measuring techniques in physics. His work has been published in ca. 250 papers in international journals and has led to ca. 350 presentations at international conferences. His current research interests include electromagnetic theory, computational electromagnetics, planar antennas and circuits, nano-electromagnetics, EM radiation, EMC, and bioelectromagnetics.

Dr. Vandenbosch was a member of the Board of FITCE Belgium, the Belgian branch of the Federation of Telecommunications Engineers of the European Union, from 2008 to 2014. From 2001 to 2007, he was the President of SITEL, the Belgian Society of Engineers in Telecommunication and Electronics. From 1999 to 2004, he was the Vice-Chairman, and from 2005 to 2009, he was the Secretary of the IEEE Benelux Chapter on Antennas en Propagation. He holds the position of the Chairman of this Chapter. From 2002 to 2004, he was the Secretary of the IEEE Benelux Chapter on EMC. From 2012 to 2014, he was the Secretary of the Belgian National Committee for Radio-electricity, where he is in charge of Commission E.

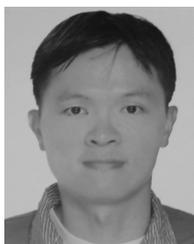

**Cheng-Ming Chen** (S'16) received the M.S. degree from the Graduate Institute of Communication Engineering, National Taiwan University, Taipei, Taiwan, in 2006. He is currently pursuing the Ph.D. degree with the Katholieke Universiteit Leuven, Leuven, Belgium, investigating distributed massive MIMO system with software defined radio.

From 2006 to 2011, he was with the Industrial Technology of Research Institute (ITRI), Hsinchu, Taiwan, where he was involved in the baseband design of WiMAX and LTE. He also involved in the 802.16m standardization with ITRI. From 2011 to 2015, he was with Broadcom Corporation, Irvine, CA, USA, as a Senior System Design Engineer, mainly focused on WiFi receiver performance verification, which include received signal strength indication, sensitivity, IQ imbalance, and adjacent channel interference. His research interests include signal processing, MIMO, coding theory, and optimization.

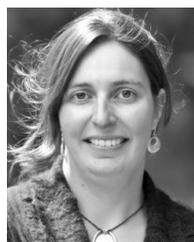

**Sofie Pollin** (S'02–M'06–SM'13) received the Ph.D. degree (Hons.) from the Katholieke Universiteit Leuven, Leuven, Belgium, in 2006.

From 2006 to 2008, she continued her research on wireless communication, energy efficient networks, cross-layer design, coexistence, and cognitive radio with the University of California at Berkeley, Berkeley, CA, USA. In 2008, she returned to imec, Leuven, to become a Principal Scientist with the Green Radio Team. Since 2012, she has been a tenure track Assistant Professor with the Electrical Engineering Department, Katholieke Universiteit Leuven. Her current research interests include networked systems that require networks that are ever more dense, heterogeneous, battery powered, and spectrum constrained.

Dr. Pollin is a Belgian American Education Foundation and a Marie Curie Fellow.

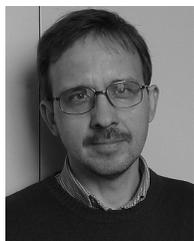

**Vladimir Volski** (M'00) graduated from and then received the Ph.D. degree from the Moscow Power Engineering Institute, Moscow, Russia, in 1987 and 1993, respectively.

In 1987, he joined the Division Antennas and Propagation of Radio waves, Moscow Power Engineering Institute, as a Researcher. Since 1996 he has been a Researcher with the ESAT-TELEMIC Division, Katholieke Universiteit Leuven, Leuven, Belgium. His current research interests include electromagnetic theory, computational electromagnetics, antenna design, and measuring of electromagnetic radiation.